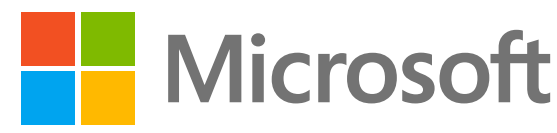

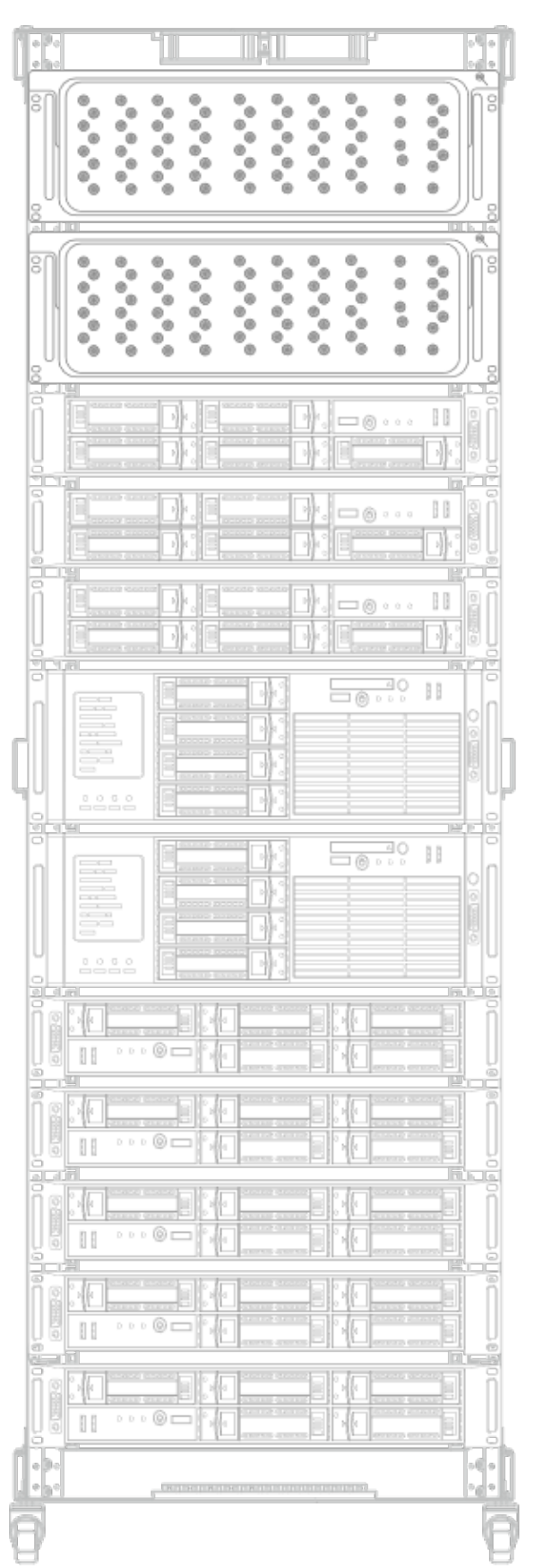

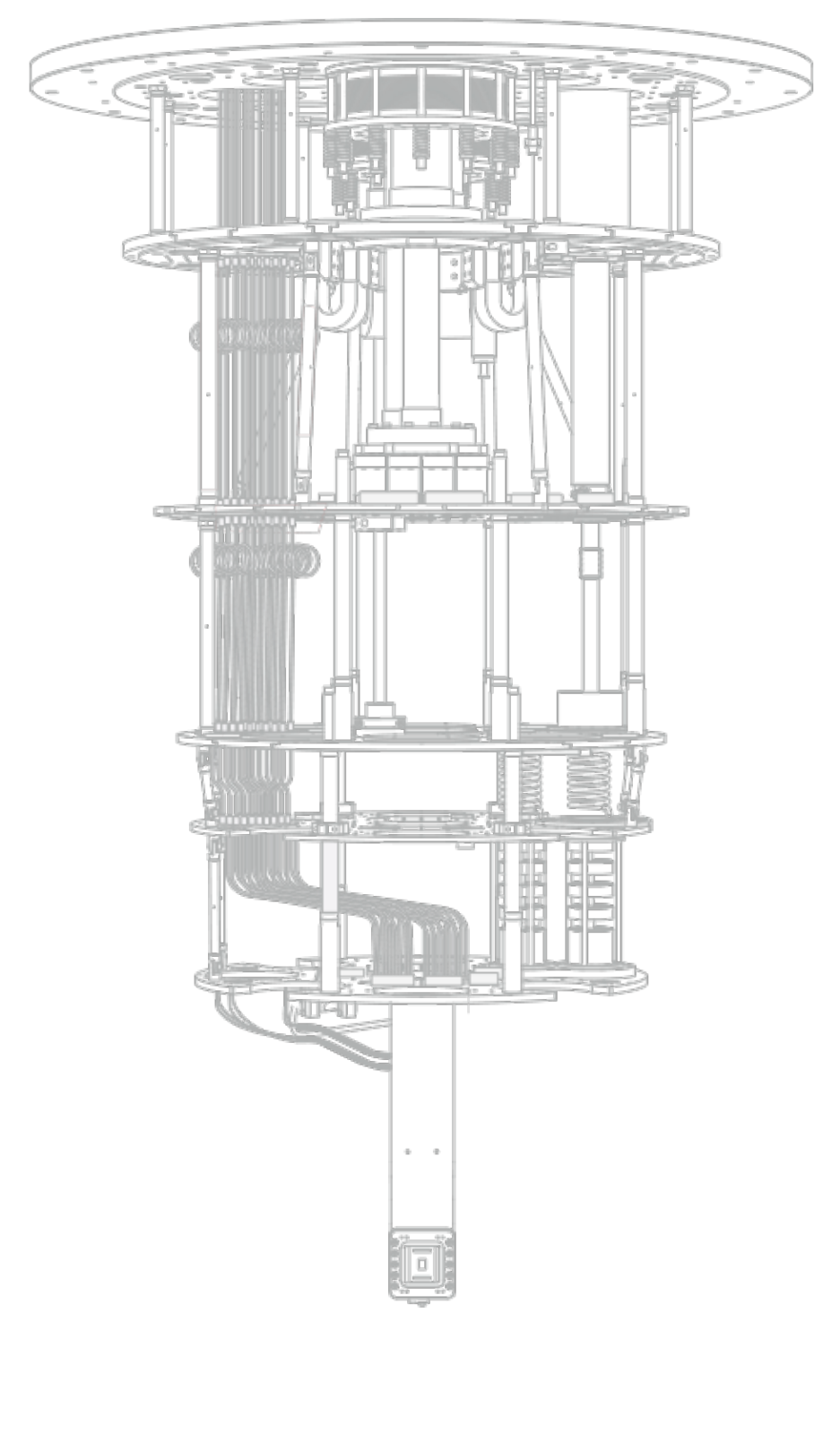

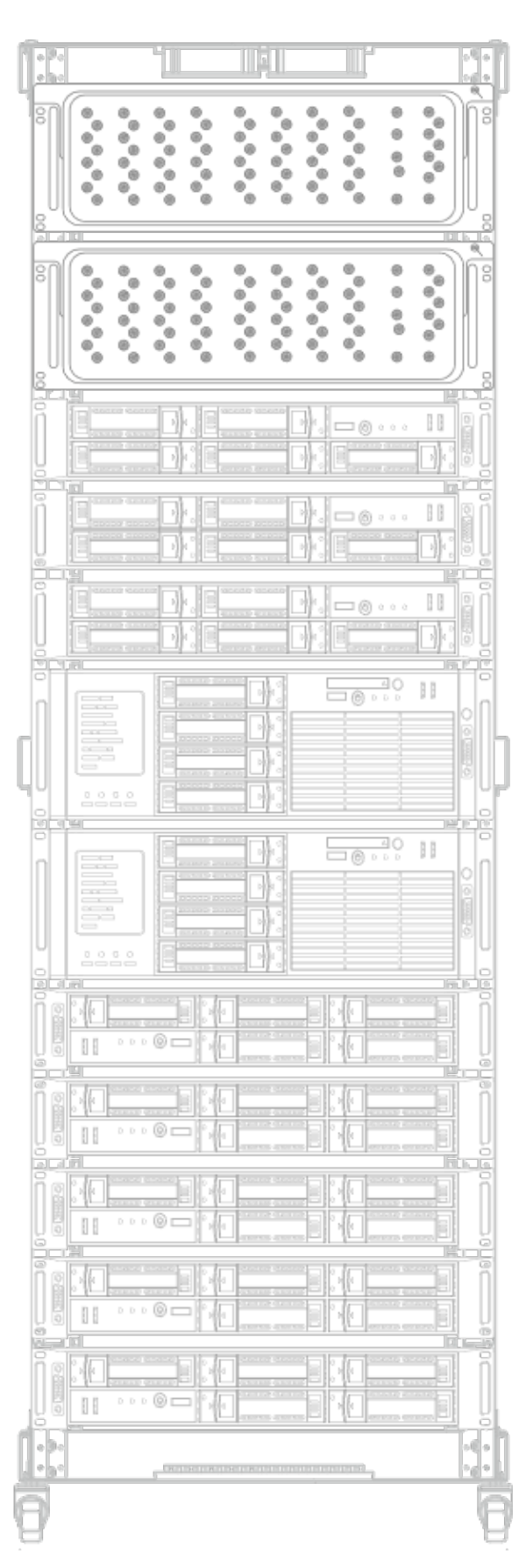

# Quantum computers will not be that different

## A blueprint for quantum computer architecture at scale

Torsten Hoefler, Microsoft and ETH Zürich
Matthias Troyer, Microsoft

# Table of Contents

# Abstract

Quantum computers are technologically novel and unusual, but at system scale they should be engineered using many of the same principles that govern classical heterogeneous accelerators. This paper argues that utility-scale quantum architecture is primarily a cost-performance problem across a coupled quantum-classical system, leading to a blueprint for scalable quantum processing unit (QPU) design. Our architecture blueprint is organized around clean logical abstractions, hiding details and complexity of physical qubit modalities below the instruction set architecture (ISA) boundary, and specializes recurring functions aggressively to minimize the cost for utility scale quantum computations. Its low-level implementation through specialized local hardware for control, readout, and quantum error correction (QEC) closely resembles the architecture of high-performance network stacks. One of our main insights is that the design principles and the resulting architecture closely follow established practice from classical computing and networking.

### Key insights and principles

1. Utility scale should be defined by economically meaningful applications, time to solution, and cost per quantum operation, rather than by qubit count alone.
2. At the system level, a quantum computer is best treated as a cloud-attached heterogeneous accelerator, so scheduling, orchestration, resource management, and integration can largely reuse classical high-performance computing (HPC) and cloud principles.
3. QEC, control, and readout should be implemented in hardened local hardware, analogous to error correction and signal processing in high-speed networking and wireless physical layers (PHYs), rather than exposed as ordinary software workloads.
4. The architectural stack should preserve familiar abstraction boundaries: compilers target a portable quantum ISA, while lower layers translate logical operations into code-specific micro-operations, physical controls, measurements, and decoding.
5. Functional specialization must be a guiding principle: the design and implementation of compute regions, memories, magic-state factories, communication interfaces, data loaders, and future arithmetic units should be optimized separately.
6. Abstraction placement and physical locality are first-order architectural decisions because bandwidth, latency, wiring, power, cooling, and signal integrity determine where control, readout, and decoding can realistically be performed.

# 1 Introduction

Quantum computing is approaching utility scale, the point where practical applications are becoming plausible, which makes questions of architecture increasingly urgent. Recent work in quantum computer architecture has already begun to converge on a view of quantum systems as accelerators attached to classical computing environments rather than as entirely standalone machines [1]. In this paper, we argue that this convergence is not accidental: cost, performance, and integration requirements will drive quantum computer architecture toward many of the same tradeoffs that shaped classical computer design, albeit with important quantum-specific constraints. Building on that perspective, we outline a blueprint for large-scale quantum computer architecture that is intended in the same spirit as the early von Neumann and Harvard models [2,3] not as a fixed implementation, but as a guiding framework for the coming decades.

We therefore organize the discussion by covering the full stack, from cloud and system architecture to ultimately Quantum Processing Unit (QPU) microarchitecture. A central claim of this paper is that the distinctly quantum aspects of architecture are concentrated largely in the lower layers, where qubit control, measurement, error correction, cryogenic constraints, and specialized physical interfaces dominate the design space. At higher levels of abstraction, however, the picture is much more familiar: quantum systems look like classical accelerators embedded in larger computing environments, with many of the same architectural questions around scheduling, resource allocation and management, programmability, networking, and integration into heterogeneous cloud and HPC systems. In that sense, much of the upper stack is not merely analogous to classical accelerator architecture [4], but in many respects effectively identical to it.

That lower stack is dominated first and foremost by the need for Quantum Error Correction (QEC) [5]. Unlike classical bits, which can be stored and processed in transistors with error rates on the order of $10^{-18}$ per cycle and are therefore effectively reliable at the architectural level, physical qubits typically exhibit error rates around $10^{-4}$ for both storage and computation. On top of the higher error rates, one needs to keep in mind that quantum information cannot be copied, and hence widely used checkpoint and restart approaches are not feasible. Useful quantum machines therefore must continuously correct errors while the computation is in flight. This makes QEC fundamentally different from most classical protection schemes, which primarily protect memory or communication; in quantum computing, error correction must also protect the computation itself, because operations propagate and transform errors and the decoding process must evolve with the executed program rather than remain static. At the same time, QEC is not a monolithic function. Codes and decoding flows can be tuned to specific circuits and operations, which opens the door to hardened implementations with specialized signaling, connectivity, and control paths tailored to particular codes and functions. Here the similarity to established classical approaches to error correction in *high-performance networking* seems striking.

Throughout our analysis a broader architectural principle emerges: large-scale quantum computers will need functional specialization just as classical computers do.

In classical architecture, substantial efficiency gains come from hardening frequently used functions into dedicated structures rather than implementing everything in a fully general reconfigurable substrate. Arithmetic units, memory systems, and input/output (I/O) paths are optimized separately because this yields better cost, speed, and power; indeed, hardened application-specific integrated circuit (ASIC) implementations can achieve up to 40× lower area, 4× higher speed, and up to 12× lower power than comparable lookup-table-based reconfigurable designs [6,7]. The same logic applies in quantum computer architecture. If general purpose quantum registers, quantum memory, magic state preparation, communication, and in the future potentially also arithmetic are all treated as fully generic functions, the resulting machine will be unnecessarily expensive and inefficient. Instead, significant gains can be achieved by constraining functionality where possible and implementing recurring operations in specialized hardened units, plausibly yielding more than an order-of-magnitude improvement for important Quantum Computer Architecture functions.

The same need for specialization is particularly clear for the massive classical computation required by quantum error correction. That processing should *not* be viewed as conventional software running on general-purpose processors. A closer analogy is the specialized error-correction and signal-processing hardware found in high-speed wireless and networking PHYs, where enormous numbers of operations are executed close to the physical interface in hardened circuits rather than exposed as ordinary compute. A 200G Ethernet PHY, for example, can require on the order of 100 tera-operations per second for signal processing and error correction [8]. Yet such work is not treated as general-purpose computation because it is implemented in specialized local hardware. Quantum systems will require the same architectural move: keep DAC/ADC, control, readout, and QEC workloads close to the physical interface, specialize them aggressively, and avoid exporting them as ordinary software-visible computation. The detailed implications for abstraction placement and locality are taken up in the next section.

As a result, practical quantum computer architectures are likely to evolve toward collections of specialized components — quantum memories [9], resource state factories [10,11], data loaders [12], arithmetic units such as quantum adders [13], communication interfaces [14], and other fixed-function blocks — each co-designed with the codes, control paths, and microarchitectural support best suited to its role.

In this paper, we apply an engineering mindset to the definition of a generic quantum computer architecture, with the goal of providing a framework that can both guide the design of concrete systems and support the evaluation of alternative implementations. Rather than starting from what is possible in principle alone, we adopt a cost-constrained methodology, because practical architecture will ultimately be shaped as much by economic limits as by physical feasibility. Our discussion is largely independent of the underlying qubit technology (whether based on electrons, atoms, photons, or other modalities) although some of the assumptions we make may require substantial breakthroughs for specific platforms. Concretely, we begin by selecting target applications, derive their resource requirements, and then use those requirements to work downward through the architectural stack to obtain a blueprint that should lie within roughly an order of magnitude of any realistic large-scale design. More broadly, we intend this methodology to be useful not only for the generic blueprint developed here, but also for instantiating more specific architectures once detailed technology parameters are available. In that sense, this paper is aimed at computer scientists, architects, and engineers seeking guidance on the design of quantum computers and on their eventual integration into classical cloud and HPC environments.

# 2 Utility scale means low cost for real applications

Utility scale should be understood not merely as reaching large qubit counts, but as the point where a quantum computer can solve practically relevant problems at a cost and time-to-solution that make deployment economically meaningful. Important applications have always driven the development of computing systems, and quantum computers will be no exception. In this paper, we focus on two representative application classes that have both practical significance and well-developed resource-estimation literature: (1) Shor's algorithm [15] applied to the cryptographically relevant problems underlying widely deployed public-key systems, including both RSA and elliptic-curve cryptography; and (2) quantum chemistry calculations [16] that go beyond what is classically tractable, such as strongly correlated active-space problems on the order of roughly 50 orbitals. These examples let us anchor the architectural discussion in concrete utility targets rather than abstract asymptotic possibilities, and they provide a useful span from cybersecurity applications to scientifically and industrially relevant simulation problems.

While there are tools like the Microsoft Quantum Resource Estimator [17] that can provide accurate estimates for required resources (qubits and runtime) based on a concrete implementation of a quantum algorithm, compilation, and detailed architecture model for the quantum computer, we work with order of magnitude estimates. Embracing specialization we distinguish between general purpose compute qubits, more efficient memory qubits with limited operations that have been the focus of recent optimizations [18], and so-called magic state factories that produce resource states for Toffoli or rotation (T) gates. Rough performance models for these applications give surprisingly similar numbers for both applications. Breaking RSA-2048 requires about 130 compute qubits, 1200 memory qubits and magic states for 1 billion Toffoli with error rates less than $10^{-13}$. For breaking elliptic curve cryptography, we require a similar number of qubits but only about 100 million Toffoli gates [19]. For our chemistry examples, we need about 100 compute qubits, 1000 memory qubits and 50 million to one billion magic states [20]. In general, runtimes typically scale with the number of magic states, because a small number of logical cycles, typically a single-digit number, is needed to consume each magic state [21].

Because physical qubits are noisy, the application-level error rates required for utility-scale computations cannot be achieved with bare qubits alone. They require QEC, which encodes each logical qubit into many physical qubits and continuously detects and corrects errors during execution. This protection comes with substantial overhead in both space and time: each logical qubit must be represented by a set of physical qubits and every logical operation expands into many physical cycles of syndrome extraction, decoding, and correction. For the order-of-magnitude estimates in this paper, we therefore assume a machine with roughly 1 million physical qubits operating at a physical cycle time of about 1 μs and logical cycle time per fault tolerant operation of 10 μs. These are strawman numbers and the assumptions can be modified by the reader and be replaced by more accurate modeling. Depending on the qubit architecture, QEC code, and architecture details, reasonable designs could range from a few hundred thousand to a few million physical qubits, with physical cycle times ranging from about 100ns [22] to 100μs or 1ms [23], resulting in logical cycle times from under 10μs to 10ms and longer. We choose numbers on the more optimistic and thus more demanding side for our analysis.

These physical qubits do not exist in isolation. Each one must be embedded in a classical control and readout stack that interprets the program representation, issues the required control signals, triggers measurements, and participates in error correction. The exact signal modality depends on the qubit technology, be it solid state qubits, atoms, ions, photonic qubits. Control signals can be a mix of  voltage, current, magnetic flux, radio-frequency excitation, light, or other analog controls—but in every case the architecture requires digital-to-analog conversion (DAC) and associated analog drive hardware to control the qubit, together with analog measurement hardware and analog-to-digital conversion (ADC) to read it out. Any realistic quantum computer architecture is therefore unavoidably a complex combined quantum-classical system, in which the qubit array is paired with substantial classical infrastructure for control, sensing, and low-latency processing.

These assumptions let us derive a simple cost envelope for utility-scale deployment and what is needed to achieve it. Let us do a back of the envelope estimate: If a quantum computer can be offered at roughly $1,000 per hour, then – with the above assumptions on physical clock speed on the order of 1μs and logical clock speed on the order of 10μs many target applications would fall into a total project-cost range of $10,000s per job execution, assuming tens of hours per job, and then taking  $100,000 to a few million $ for a project, for a which is comparable to present-day resource commitments for major supercomputing efforts.

At $1,000 per hour, continuous operation corresponds to about $9 million per year, or roughly $45 million over a five-year lifetime for capital and operations combined. If we assign about 40% of that total to operations and 60% to capital investment, the available capital budget is about $27 million, which in turn suggests that the quantum computer itself should cost on the order of $20 million, reserving $7 million for surrounding infrastructure. For a machine with about 1 million physical qubits, that implies a target cost of only around $20 per qubit including the integrated quantum and control stack, underscoring how central control integration and architectural specialization will be to practical system design.

Note here that slower clock speeds directly translate into longer runtimes and thus higher cost, requiring a commensurate reduction in cost per qubit. For example, all else being equal, ten times longer cycle times imply ten times longer runtimes and a desired ten-fold reduction of the cost per qubit to $2.

Given these observations, reaching utility scale at low cost is not mainly a question of maximizing qubit count, but of managing the architecture so that complexity, bandwidth, latency, and hardware specialization remain economically viable across the full stack. The key design challenge is therefore, similarly to classical computer architecture, to identify which functions should remain programmable and which should be hardened, where abstractions should hide lower-level implementation details to simplify processing, and how tightly classical and quantum components should be integrated to minimize both system cost and time to solution.

The main pillars that emerge for Quantum Computer Architecture design are:

1. **Heterogeneous system architecture:** Designing the quantum cloud computer like a classical heterogeneous system, so scheduling, orchestration, and resource management can reuse well-understood cloud and High-Performance Computing (HPC) principles. In that sense, a QPU should be treated like a cloud-based accelerator even though first instantiations are likely to live in separate racks.
2. **Functional specialization:** Use specialization aggressively, including at the ISA and microarchitectural levels, because fixed-function or constrained-function units can reduce cost, bandwidth, and control complexity relative to a fully generic design. Specifically, utilize specialized functional units—such as compute, memory, magic state factories, and communication units—since this structure clarifies both the ISA requirements and the opportunities for hardened implementations.
3. **Abstraction boundaries:** Maintain clean abstraction boundaries, especially between the logical ISA and the underlying QEC implementation, so tradeoffs in code choice, bandwidth, and hardware realization can be evaluated and optimized without exposing unnecessary physical detail. Pick those abstractions to balance the benefits of co-locating classical and quantum functions against the cost, cooling, wiring, and bandwidth penalties of tighter integration.

A note on terminology: some people call quantum programs “circuits” but this is deceiving – in CS/EE a circuit always has all elements active and can usually pipeline data. A quantum program is really a sequence of operations without pipelining. We avoid the terms circuit and gates to avoid this confusion.

# 3 Heterogeneous System Architecture: From QPUs to Clouds

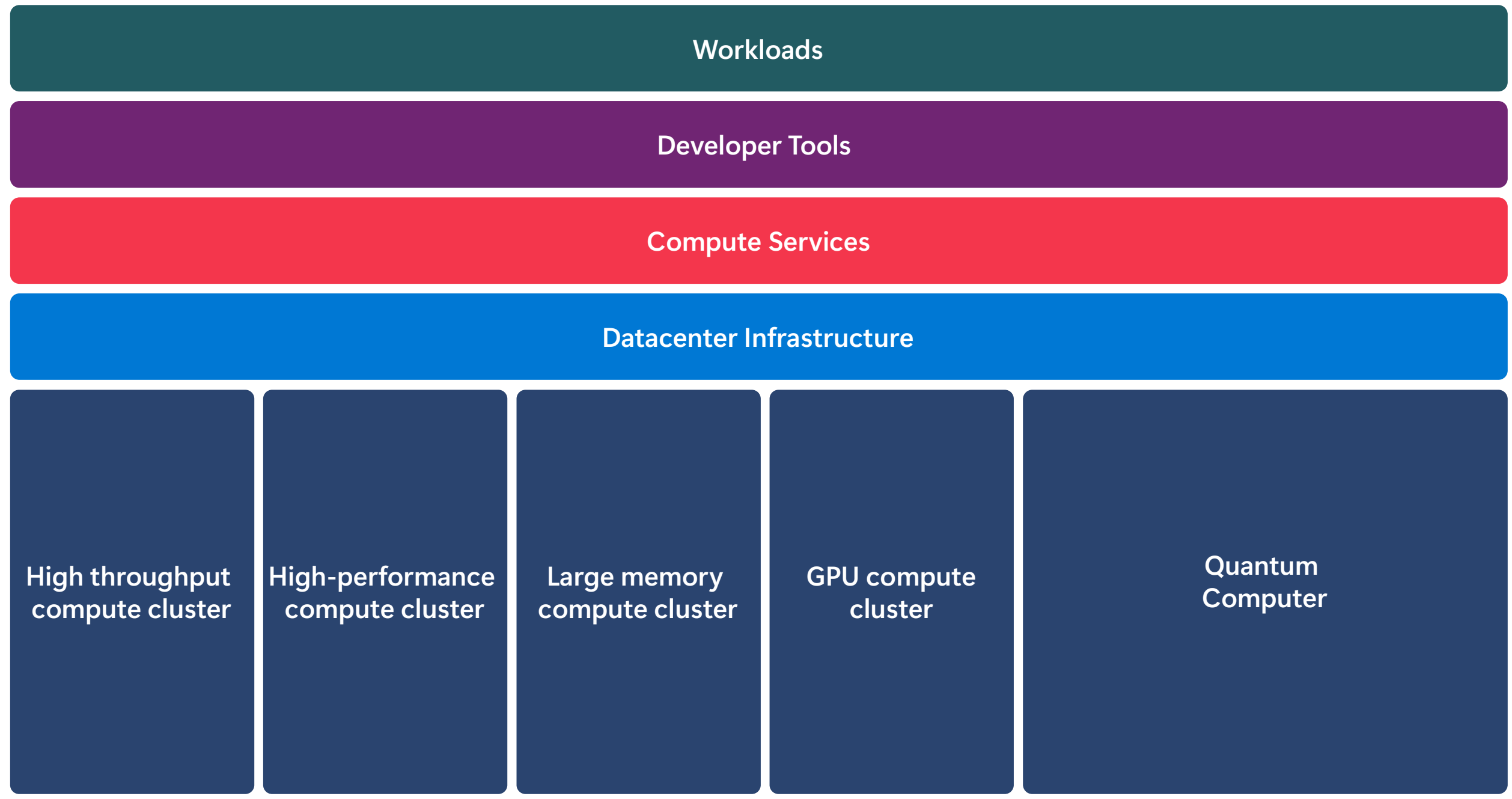

*Figure 1: At cloud scale, a quantum computer fits into the same architectural pattern as other nodes and clusters, including accelerator based ones, from user workloads, and developer tools down to compute services, datacenter infrastructure and compute hardware . A quantum node consists of a classical frontend machine and a quantum accelerator, and multiple such nodes can be deployed, scheduled, and managed using established datacenter principles used for heterogeneous systems in HPC and cloud. The quantum-specific complexity is largely internal to the accelerator; most scaling across nodes relies only on standard classical networking, with a quantum backend required only when multiple QPUs must be combined into one larger logical machine.*

While qubits enable fundamentally different computation inside the QPU, the system architecture above that level is not radically different from classical HPC or heterogeneous computing systems. A quantum computer is best understood as a classical computer equipped with a QPU, much as modern systems are built around CPUs coupled to GPUs or other accelerators. It thus fits well into the general architecture from user workloads to datacenter hardware as shown in Figure 1. This similarity substantially simplifies integration into cloud and heterogeneous compute environments, because many of the same principles for orchestration, scheduling, and system composition apply – nothing quantum specific is required.

Heterogeneous workloads use many of the same developer tools, software engineering practices, and runtime models as purely classical codes. Concepts from HPC such as MPI [14,24], optimized libraries [25], workflow systems, and auto-tuning [26] carry over naturally to quantum computing. Much of the distinctiveness therefore lies not in the upper software stack, but in the specialized units below, which we discuss next. Higher-level toolchains will increasingly hide that complexity, and AI-assisted development is likely to accelerate this trend; we expect that over time, most quantum code will be written by AI.

Even the network is similar: standard classical frontend networks are sufficient in most cases, while a specialized quantum backend network is only required when a single quantum system does not provide enough qubits or enough performance to hold or execute a target application within the required time.

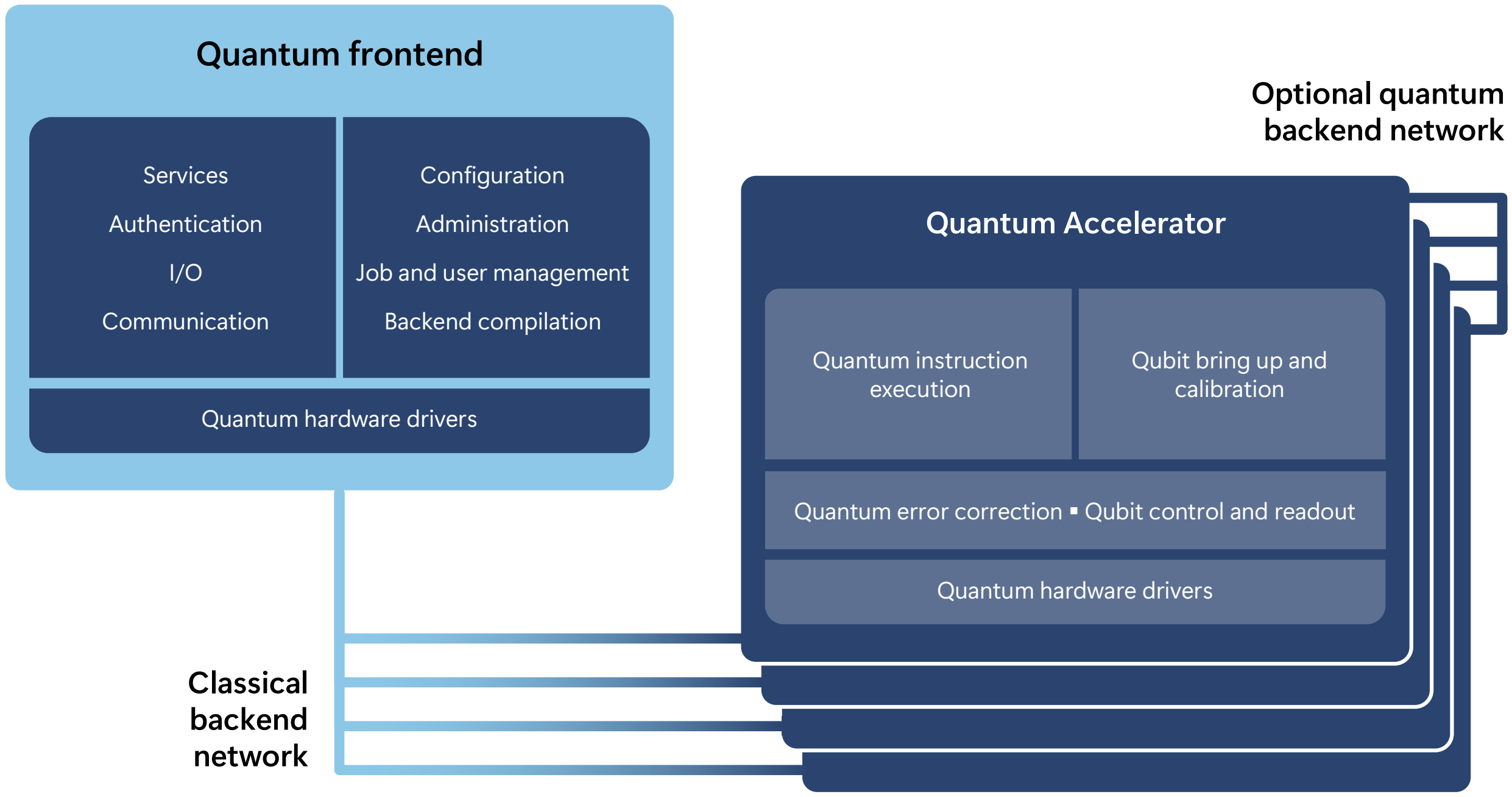

*Figure 2: Components of a quantum compute cluster. A quantum cluster consists of frontend machines for job submission and system management, quantum accelerators with tightly coupled subsystems for bring-up and calibration, instruction execution, and control/readout/QEC, and networking that ranges from standard classical datacenter interconnects for throughput scaling to a quantum backend only when multiple QPUs must be combined into one larger logical machine.*

Coming to the quantum compute cluster itself, we see the need for the following components:

- **Frontend machine.** Quantum computers will be connected to a frontend machine responsible for system management. Architecturally, this machine plays a role similar to a host CPU or service node in classical accelerator-based systems: it accepts user jobs, handles authentication, scheduling, and resource management, handles backend compilation, stages program and data movement, manages classical input and output, and coordinates execution on the quantum accelerator. It is also the natural place for workflow integration with the surrounding cloud or HPC environment, because most user-visible interaction with the system remains classical even when the main computation is quantum. In that sense, the frontend machine is not part of the quantum accelerator itself but the system component that makes the quantum accelerator usable as a shared, managed cloud resource.

- The **quantum accelerator** can be understood as a stack of tightly coupled subsystems:
  a. A **bring-up and calibration subsystem,** analogous in spirit to BIOS or power-on self-test in classical systems, but much more complex. It characterizes the device including the qubits, establishes operating points, saves calibration data, and verifies that the hardware is ready for reliable execution. Re-calibration of qubits may be required between job executions or for long-running jobs even during execution.
  b. The **instruction-execution subsystem,** which carries out the quantum program on a combination of general-purpose compute qubits and, where available, specialized units such as memories, magic-state factories, communication interfaces, and other future specialized units.
  c. At a lower level in the stack there is the **control, readout, and error-correction subsystem.** It is implemented through tightly coupled specialized classical processing that translates instructions into control signals, then demultiplexes and interprets measurement data, and performs the continuous decoding and feedback required for quantum error correction. In aggregate, this lower subsystem can require petascale-class classical processing, but like a networking PHY or accelerator control plane it should be viewed as hardened local infrastructure rather than as ordinary software-visible compute.
  d. And finally, the **qubit** plane that houses the qubits, physical representation of the quantum states.

These subsystems are coordinated by a **quantum accelerator firmware,** analogous to the firmware used in GPU and network accelerators, which is not visible or accessible to the users and manages bring-up, calibration, and the control interface between the frontend machine and the quantum hardware.

- **Networking and cluster computing.** Scaling to multiple quantum computers in a cloud can proceed in two distinct ways.

  a. The first, and likely the most common, is a classical cluster model in which many independent quantum computers are deployed as accelerator nodes in a datacenter and connected through ordinary frontend classical networks. In this mode, the goal is higher throughput, greater availability, and multi-tenant cloud operation rather than a single distributed quantum computation. Because quantum clock rates are relatively low and user visible I/O remains modest, these systems do not require fast classical backend networks.

b. The second mode is a distributed-quantum model in which multiple QPUs are linked by a quantum backend network that can establish remote entanglement and teleport logical states between QPUs. This is only needed when a single QPU does not provide enough qubits to hold the application or enough performance to complete it within the required time.

In other words, classical networking suffices for scaling out cloud capacity, while a quantum backend is reserved for scaling up the effective size or speed of one logical quantum computer. A quantum frontend network could be added in the future for quantum communication with quantum sensors or distant quantum computers [27].

# 4 Functional Specialization: Towards performant QPU architecture

Large-scale quantum computers will – at least initially – include **generic logical qubits** as a fully programmable and reconfigurable compute fabric. They are the only qubits available on first machines. These qubits must support a broad set of logical operations together with measurements and flexible data movement patterns needed to implement arbitrary algorithms.

Beyond them, specialization is needed to contain cost, much as in classical computing, where one could imagine a fully reconfigurable machine but in practice relies on specialization to make computation efficient and affordable. Classical systems combine general-purpose computation with specialization at all levels: memory is distinct from CPUs to save cost. CPUs contain optimized ALUs, are paired with specialized accelerators such as GPUs and NPUs, and connect to other devices using specialized data data-transmission circuits.

The same principle applies in quantum computing: by constraining functionality to a specific set of operations, one can design hardened implementations that are substantially cheaper than fully general-purpose compute qubits. Representative savings of 3× for quantum registers, and potentially an order of magnitude for memory, magic-state factories, and quantum networking units can be realized. In architectural terms, this means that a scalable quantum computer should not treat all logical qubits as identical.

Quantum specialization has several closely related benefits. First, fixed-function units can use specialized QEC schemes that reduce qubit overhead relative to fully general compute regions. Second, they can simplify the underlying physical qubits and microarchitecture by restricting a supported operation set—for example, only a magic-state factory may need rotation gates. Third, specialization enables optimized hardened implementations of control, connectivity, and local data movement tailored to the unit’s role, such as limited functionality for memory qubits.

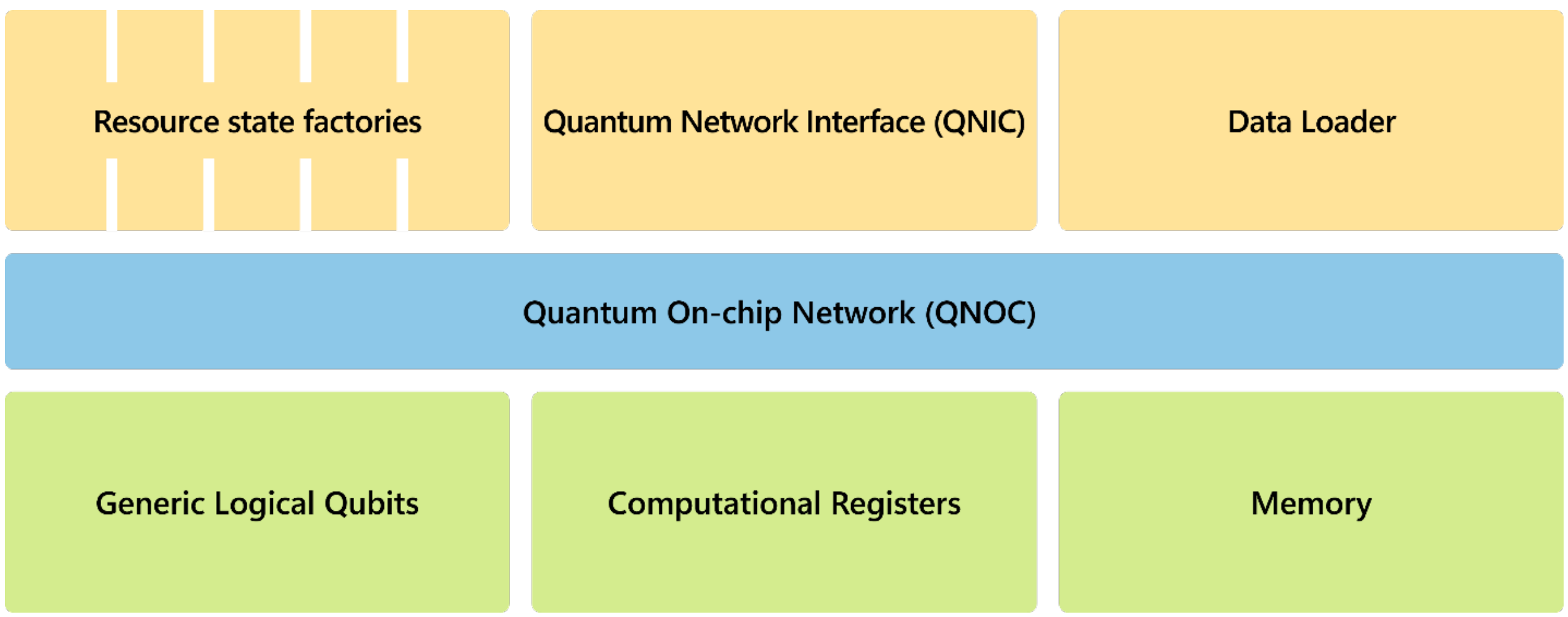

*Figure 3: QPU layout example for a quantum chip illustrating functional specialization combined with generic quantum compute and interconnection.*

Examples of specialized units include quantum memory and computational quantum registers, resource state factories, quantum network interfaces, data loaders or state-preparation blocks, and in the future potentially quantum ALUs. We discuss each of these in more detail below.

**Resource state factories** are a common distinct architectural unit because they supply the non-Clifford resource needed for universal fault-tolerant computation. Clifford operations such as H, S, CNOT, preparation, and measurement are not sufficient on their own. To go beyond them, architectures typically prepare ancilla states such as $|T\rangle$ or $|CCZ\rangle$ and consume them through gate teleportation or injection [10,21]. Beyond their essential role in non-Clifford gates, resource states can also be used to accelerate Clifford gates or QEC syndrome extractions [11,28].

Non-Clifford states are usually prepared starting with many noisy candidates and purifying them through magic state distillation or cultivation, a sequence of Clifford-only verification and post-selection steps that produces fewer, higher-quality outputs [10]. In practice this forms a staged pipeline in which changing the code or code distance is itself an important form of specialization and optimization: early stages can use smaller-distance, cheaper codes, while later stages increase the code distance only once the surviving states are valuable enough to justify stronger protection. More broadly, resource state factories can employ specialized codes and layouts, further reducing overhead. This can lead to an order of magnitude reduction of physical qubits.

Once a quantum computer contains magic state factories, all other physical qubits can be simplified to only perform Clifford gates.

**Quantum memory** or **quantum SRAM (static random access memory)** deserves to be treated as a distinct architectural unit because storing logical quantum states is a different problem from performing fully general logical computation on them. General-purpose logical registers need to balance qubit encoding overhead (QEC encoding rate) and efficiency of operations, which tends to favor QEC codes with few logical qubits per block. A dedicated quantum memory, by contrast, can expose a much narrower interface—primarily storage plus teleportation-based retrieval or movement—which allows the underlying QEC code to be optimized for encoding and save an order of magnitude physical qubit count with high-rate QLDPC codes [29].

**Computational quantum registers** with limited instruction sets are another opportunity for specialization and optimization. For example, yoked surface codes reduce the qubit overhead can be reduced to around one-third of that of standard surface-codes [9]. These codes can be used for more than memory. By offering a limited instruction set going beyond storage and retrieval of quantum information, they can, for example, be used to implement time evolution of local quantum models approximately three times cheaper than general purpose qubits.

**A quantum network interface (QNIC)** is the specialized communication unit of a quantum computer [14]. Its purpose is to move logical quantum states between QPUs using remote entanglement to facilitate quantum teleportation. In architectural terms, it is the quantum analogue of a classical NIC: a hybrid quantum-classical interface optimized for reliable communication rather than general computation, managing entanglement resources, physical link operations, and a conventional classical coordination network.

Quantum teleportation requires entangled Bell pairs between sender and receiver and a classical network to send two classical bits of information. The qNIC must continuously replenish high-fidelity entanglement through remote entanglement generation followed by entanglement distillation, and it must couple this entanglement-management pipeline to the local teleportation and control logic. Despite the quantum nature of the transfer, the classical side-channel requirements remain modest: even for an extreme upper bound in which a 10,000-qubit machine teleports roughly 5,000 qubits per 10 μs cycle would require only about 1 Gbit/s of classical bandwidth for the two-bit teleportation outcomes, which is well within the range of modern classical interconnects.

The physical qubits in the QNIC need to be specialized for entanglement generation through a quantum network. The logical qubits can be optimized for efficient entanglement distillation: similar to magic state distillation this forms a staged pipeline in which early stages can use smaller-distance, cheaper codes, while later stages again increase the code distance saving a factor of 2-3x.

**Quantum loaders, ROM (read-only memory) units, or state-preparation blocks** are another natural class of specialized unit. Their role is to initialize a set of qubits from classical data. Because this task is more constrained than fully general computation, it also opens an interesting optimization space: there exist algorithms for data loading where the required fidelity scales sub-linearly with the data size and number of operations. This and the use resource states suggests that loaders may gain efficiency both through smaller distance codes and reduced control complexity [12].

**Quantum ALUs (arithmetic logic unit)** are specialized units for optimized structured computation, just as classical ALUs use specialized circuits for fixed functions such as addition or related arithmetic. The challenge is that reversibility makes such units less straightforward than their classical counterparts: a generic quantum adder typically requires a substantial set of ancilla qubits, whereas in a larger circuit those ancillas can often be absorbed into surrounding structure and reused, or specialized faster un-compute operations can be employed. This creates a nontrivial tradeoff between the convenience and speed of a dedicated arithmetic block and the overhead of reserving additional qubits for it.

**"On-chip" quantum networks (QNOC)** finally connect these specialized functional units. In classical systems, one would define a data path made of wires, switches, and repeaters to move data between blocks. Quantum architectures require an analogous mechanism for moving qubits and translating ("switching") between different QEC codes. QNOCs could be implemented by special purpose hardware connections like (1) wires or optical fibers, (2) physical movement of qubits, or (3) data paths built from generic or specialized qubits for "on-chip" quantum teleportation. In addition, the QNOC may use specialized qubits for code translation where required. In that sense, a reconfigurable generic compute array could serve not only as a general-purpose execution substrate but – in the absence of specialized connections – also as the interconnect that ties specialized units into a coherent architecture.

# 5 Abstraction Boundaries: Exploiting locality in QPU microarchitecture

We now focus on where to place the abstraction boundaries between the logical quantum instruction stream and the control, readout, and error-correction machinery below it, because those boundaries determine whether bandwidth, latency, wiring, power, and thermal constraints remain feasible at scale. The central idea is that the lower layers of a QPU should be organized around locality, with only compact logical instructions and results exposed upward while the overwhelming volume of device-specific activity is absorbed by specialized hardware.

This is **analogous to a network PHY,** where a compact interface expands through local stages into encoding, lane distribution, waveform generation, analog signaling, and receive-side digitization, signal processing and error decoding while the higher-level system sees only a much smaller semantic interface. The same principle organizes control and readout of a fault-tolerant QPU: a compact quantum ISA expands on the control side ultimately to pulse generation and on the readout side through measurement processing, and QEC decoding to logical measurements. The purpose of this hierarchy is to absorb bandwidth growth and timing-critical detail in specialized local hardware rather than exposing that internal machinery as general-purpose computation.

The key architectural question is therefore the same one that arises in communication systems: which functions should remain visible and programmable at the interface, and which should be co-located, hardened, and hidden below the abstraction boundary to make the overall system scalable and economical?

## 5.1 Qubit Control

### 5.1.1 Instruction hierarchies

Against that backdrop, the first architectural consequence is that a QPU should expose a **quantum ISA** that provides an efficient way to express general quantum programs and hides the details of the underlying error-correction machinery. This abstraction plays a role similar to the boundary between an architectural ISA and the microcode or implementation mechanisms below it in classical processors: users and quantum compilers should see logical quantum instructions, not the full sequence of code-specific logical and physical operations, measurements, physical pulses, and decoding steps required to realize them. The resulting interface is best viewed as an accelerated RISC-style [30] quantum ISA: it should contain a small set of basic operations sufficient for universality, while also admitting higher-level instructions for recurring primitives such as magic states, general rotations, entanglement generation, teleportation, or cat-state creation, much as classical ISAs add vector or matrix extensions for common performance-critical patterns.

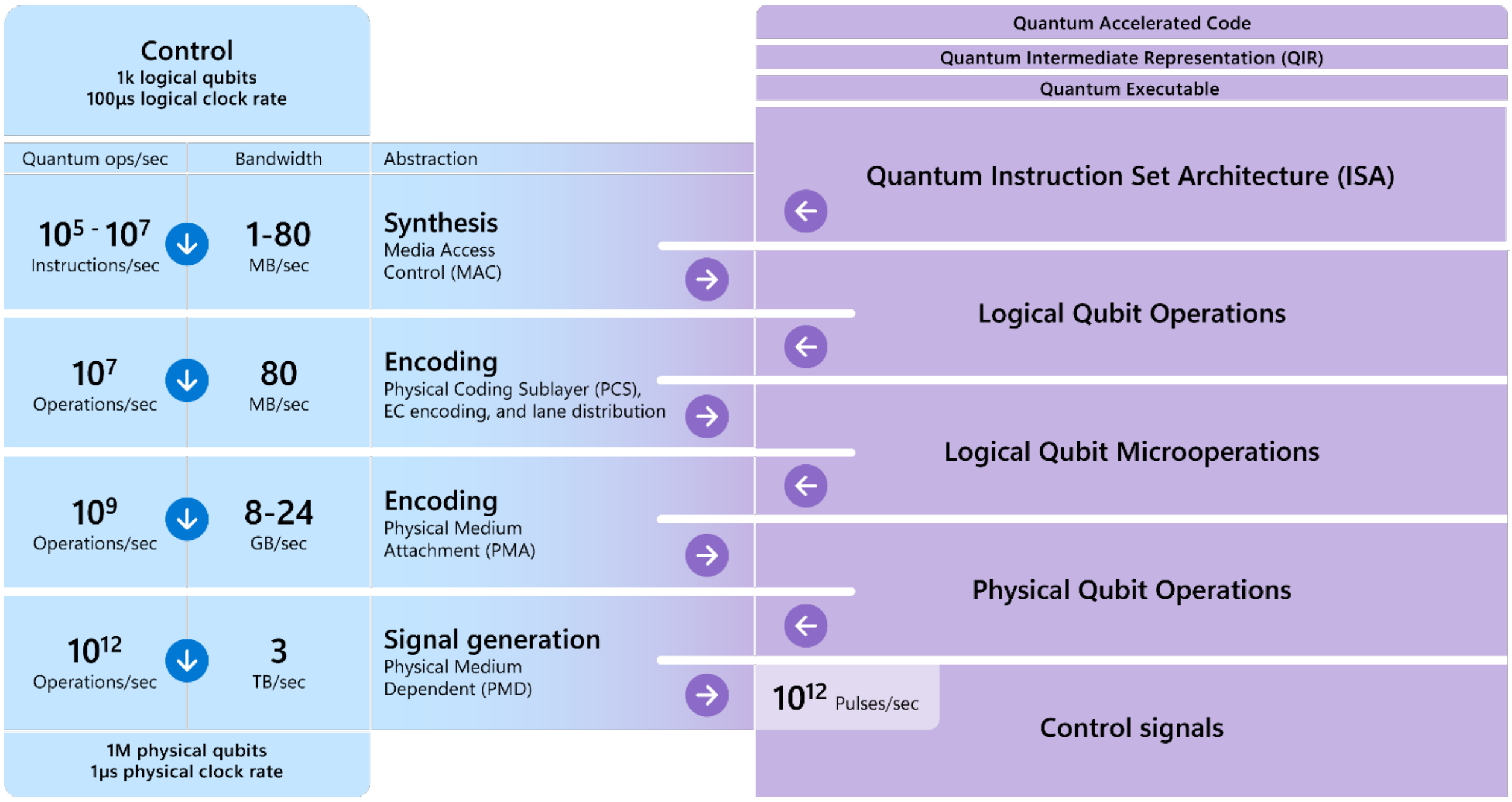

*Figure 4: Encoding hierarchy and bandwidth example for qubit control showing how required bandwidths can be reduced by local computations captured in layer-wise abstractions along the stack. This illustrates the principle of maximum locality for minimal data movement through abstraction.*

This ISA boundary is the top of a **deeper translation stack for the control path.** Starting from an assembly-level program, each logical instruction is progressively lowered into finally the physical control signals of the underlying hardware architecture [17,31], as shown in Figure 4. At the ISA level, **Logical Qubit Instructions** are visible to the quantum compiler. Following [31], these are then translated into sequences of QEC code-native **Logical Qubit Operations** (LQOs), dependent on a particular QEC code. LQOs are then further expanded into an intermediate sequence of **Logical Qubit Micro-operations** (LQMs), representing common building blocks of LQOs, such as sequences of syndrome extraction acting across a patch of physical qubits in a logical qubit. These are then mapped to **Physical Qubit Operations** (PQOs) on the up to hundreds or more physical qubits that constitute each patch. This PQO stream then finally drives the control hardware for the physical qubits, generating **physical control pulses** to manipulate the qubits. Thus, the architectural instruction stream can remain compact and portable, while the lower layers absorb the code-specific and device-specific detail needed for fault-tolerant execution.

With our PHY analogy in mind, these translation layers are easier to interpret. In classical high-speed networking, a compact MAC-level symbol is progressively translated through lane scheduling, forward-error correction, waveform generation, and finally analog signaling on the wire. In the same way, a quantum ISA should remain compact while LQOs, LQMs, and PQOs successively realize error correction, scheduling across control lanes, and ultimately pulse generation. The point of the hierarchy is the same in both cases: to absorb the bandwidth explosion and timing-critical detail in specialized local hardware, rather than exposing that internal machinery to general-purpose software.

### 5.1.2 Control bandwidth considerations and architectural placement

Where these translation stages are executed is therefore an important architectural choice rather than a purely implementation-level detail. Each boundary exposes a different bandwidth requirement and thus a different opportunity for either separation or co-location, for example in different modules or at different temperature zones in a cryostat.

For sake of simplicity in the presentation and discussion, we assume a quantum computer with 1,000 logical qubits (could be anywhere from 100 logical qubits on early machines to 10,000 in the future), a 10 kHz logical cycle (could be anywhere between 100 kHz and 1kHz), and at the physical level 1,000,000 physical qubits (could be between 100,000 on early machines to 10s of millions in the future) with 1 MHz physical cycle time (could be between 10 MHz and 10 kHz depending on the architecture), and across the stack assume an 8-bit instruction encoding at most layers. For simplicity we also do not distinguish between different QEC schemes and possible different clock speed of different components. The reader can easily adjust these numbers to the specifications of a concrete quantum architecture, where clock speeds and qubit counts could vary by an order of magnitude up or down.

At the **ISA level,** the stream is extremely compact. If an operation taking one logical cycle is executed on every logical qubit the maximum bandwidth is 80 Mbit/s but can typically be an order of magnitude lower by omitting idle operations and realizing that some instructions, like on classical CPUs, will take multiple cycles to execute.

After lowering to **LQOs,** the stream becomes denser and code-dependent, inflating each instruction to one or more code-native logical operations per logical qubits. For example, a CNOT instruction in a surface code implementation is mapped to a sequence of four single-qubit and two-qubit measurement operations, and this can be potentially up to 100 operations for some LDPC-based schemes, as the synthesis in block codes can be more complex. At this stage also idle operations get inserted to trigger syndrome extraction for quantum error correction on idle qubits. We end up with a stream of one operation per logical qubit and logical cycle with a total bandwidth of **80 Mbit/s.**

At the **LQM** level, the relevant clock becomes physical. LQMs describe operations across a patch of physical qubits. Such a patch may either be all physical qubits forming a logical qubit or a subset thereof; for example, in a surface code we may distinguish between the patch of physical qubits forming the bulk of logical qubit and boundary patches connecting them. With above assumption, this stream reaches about **8 Gbit/s** assuming one patch per logical qubit, or could, for example, be 24 Gbit/s in a surface code if we assume three patches per logical qubit.

Finally, after expansion to **PQOs** across the physical qubits in each logical patch, the bandwidth can rise to the Tbit/s scale, with one million qubits at 1 MHz physical cycle times this me be **3 Tbit/s** assuming now just 3 bits to encode one of up to eight physical qubit operations.

### 5.1.3 Architectural tradeoffs in qubit control and the challenge of quantum Rent's rule

The same logical program therefore spans many orders of magnitude in control bandwidth, from low Mb/s to multiple Tb/s as it descends the stack, which is precisely why abstraction placement matters: higher layers can be separated easily, while lower layers increasingly favor local execution close to the qubits or to the specialized control hardware that drives them. These considerations also motivate the introduction of LQMs between LQOs and PQOs, to avoid an extreme jump in bandwidth and allow for better architectural tradeoffs in placement of the translations.

Each stage of the translation and control stack can in principle be implemented far from the qubit control hardware (offering more choices) or closely co-located (reducing high bandwidth communication but potentially incurring noise challenges).

For qubit technologies that require cryogenic operation, these placement choices become additionally constrained by the thermal architecture of the system. Focusing for a moment on such cryogenic qubits there is a severe constraint for platforms operating near 100 mK. A representative budget of roughly 1 mW at 100 mK, or about 1 W at 1 K, leaves little room for per-qubit digital processing; for example, 1 million qubits driven at a 1 MHz control rate imply about $10^{12}$ control events per second, so a 1 mW budget would allow only about 1 fJ per event, comparable to the energy of a single transistor switch and therefore not a realistic target for substantial local computation. Practical systems will therefore need to address the conflicting demands of co-location and thermal limits by mixed placement strategies: higher temperatures provide more available power but increases the need to move high-bandwidth, high-fidelity control signals across thermal boundaries, where wiring density and heat leakage are tightly limited. Co-location has the opposite tradeoff: it reduces external bandwidth and improves signal quality, but only if the required classical circuitry fits within the available space and cooling budget. An often more costly alternative is splitting the quantum system into multiple QPUs each with fewer qubits.

The architectural conclusion is general: control is the central systems challenge for utility-scale quantum computing. The control path is not just a logical translation problem, because its physical implementation determines whether the required bandwidth, signal quality, and power dissipation are feasible at scale.

It is useful to frame the general placement problem in terms of **Rent's rule** [32], the empirical scaling law $T = kN^p$ that relates the number of external terminals $T$ of a subsystem to the number of internal components $N$. In mature classical integrated systems, the Rent exponent $p$ is typically well below 1 because hierarchy, locality, and on-chip communication absorb much of the connectivity. Quantum control does not naturally lie in that favorable regime [33]. A large-scale QPU looks closer to $p \approx 1$: each physical qubit usually demands its own control lines and readout resources, so the number of control connections tends to grow linearly with qubit count.

The architectural task is therefore to engineer the control and readout to mitigate that near-linear scaling through **multiplexing** and **local integration** with the exact balance depending on the qubit modality. Depending on the qubit modality, that can mean **spatial multiplexing,** such as free-space optics that addresses many qubits from shared beam-generation hardware as is used e.g. in qtomiuc ; **frequency multiplexing,** in which multiple qubits or resonators share wiring and analog front-ends while being separated spectrally; and, where the technology and energy budget permit it, **integrated local control,** which moves signal generation, routing, or demultiplexing closer to the qubits. In some implementations, **time multiplexing** can also contribute but often leads to slow-down of the clock speed.

## 5.2 Qubit readout and error decoding

Measurements are initiated by control signals and return analog measurement signals whose modality may be electrical, optical, radiofrequency, or others. In the strawman 1-million-qubit machine considered above, a 1 MHz physical measurement cycle with roughly 10 bits sampled per physical readout already corresponds to a worst-case raw ADC bandwidth of about 10 Tbit/s. Signal processing reduces this stream: if each physical measurement is represented by one or two bits, possibly including status information such as uncertain or failed readout, the post-processed physical stream is still on the order of 1–2 Tbit/s.

Only after syndrome decoding and error correction does the stream collapse from many physical measurements with measurement error rates of the order of $10^{-3}$ to $10^{-4}$ to far fewer measurements at the logical level with error rates as low as $10^{-12}$ or better. There, roughly 1,000 logical qubits measured at a 10 kHz logical rate require only about 10 Mbit/s. Similar error contracts between layers are standard in many stacks, for example, in Ethernet, the pre-FEC (PMD/PMA) bit error rate should not exceed $2.4 \times 10^{-4}$ but the post-FEC error rate leaving the PHY should not exceed $10^{-12}$. Those are lower bounds that are usually far exceeded in practice.

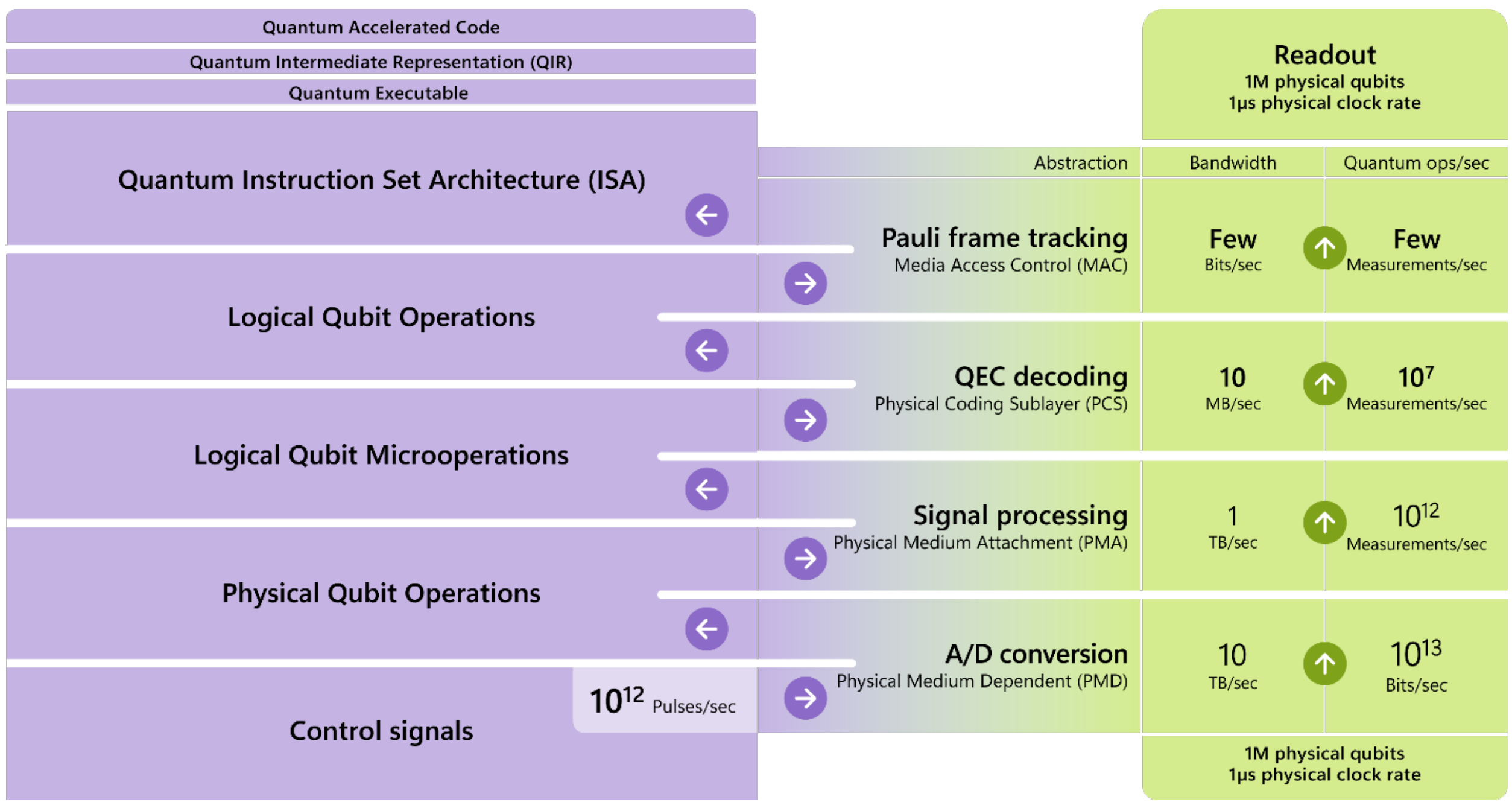

*Figure 5: Qubit measurement stack from analog signals to digital physical measurements with high error rates and then logical measurements with low error rates. Bandwidths are reduced in layer-wise abstractions progressing up the stack.*

Many of those measurements may be used in the implementation of the logical qubit instructions, as in the above example where a CNOT instruction in the surface code is implemented through a sequence of four measurements. Such measurement results returned by LQOs may be immediately used for Pauli corrections — which can be seen as another layer of error correction. Many measurements may therefore not user-visible operations at all; they are introduced by the system to perform QEC and to implement logical operations, and they should terminate before being exported through the quantum ISA. The application should see only the small number of logical measurement results that it explicitly requests, while most of the physical readout traffic is absorbed locally by specialized signal-processing, decoding, and Pauli frame-updates.

Unlike control, where qubits often require individually tuned control signals delivered to each qubit, readout can exploit substantial analog multiplexing before digitization, for example through frequency multiplexing of radio-frequency signals or spatial multiplexing with free-space optical imaging and detector arrays. This reduces the pressure from off-chip wiring and therefore mitigates or avoids Rent's rule limitations.

There are again close similarities to network PHYs. For example, weak measurement signals will often have to be amplified before substantial digital processing becomes feasible. Once digitized, demultiplexing and signal interpretation should be integrated into the ADC and readout hardware itself, much as modern Ethernet or wireless PHYs perform equalization, slicing, and decoding close to the analog interface rather than exporting raw samples to a general-purpose processor. The required computation can still be large: frequency demultiplexing with FFTs or spatial signal processing over detector arrays can reach tens of TOPS at the scale considered here; for example, demultiplexing 1,000 multiplexed values requires on the order of 5n log n, or roughly 50,000 operations, per measurement multiplexing group. At a scale of 1 million physics qubits with 1 μs measurement times this results in 50 TOPS for demultiplexing.

This kind of workload is not unusual in modern communication hardware. A 200G Ethernet PHY, for example, operates with four 53-Gbaud lanes, and equalization and signal-conditioning stages such as PAM4 slicing, CTLE, and DFE can together require on the order of 50 operations per symbol, corresponding to roughly 10.6 TOPS for four 200G Ethernet lanes [34]; concatenated Ethernet error correction adds on the order of 2 TOPS per lane [35], and projected 6G decoding workloads reach roughly 150–200 TOPS at 1 Tb/s [36]. These operations are not exposed as general-purpose computation, because they are implemented in specialized circuits co-located with the analog interface. An Ethernet 4x200G pluggable transceiver is available for around $1000-1500 in volume pricing today.

The design principles of network PHYs also apply in qubit measurement processing: quantum readout interpretation and QEC decoding are of comparable systems complexity and should likewise be realized in specialized local hardware. The same bandwidth and latency considerations then carry into QEC decoding, which should be tightly coupled to the readout signal-processing pipeline. An additional argument for specialized hardware is that latency is often the more important constraint, because measurement outcomes can determine subsequent program control flow and can limit procedures such as physical magic-state distillation. A useful design target is therefore to complete measurement, interpretation, decoding, and frame update within less than one logical cycle whenever the result is on the critical path.

# 6 Combining the Principles: Overall system design and cost

For utility scale at viable cost, the central difficulty is therefore not the qubits in isolation but the integrated control and readout system around them. Reaching the scale of roughly one million qubits at tens of dollars per qubit requires aggressive integration, whether through mixed-signal semiconductor technology, optical or acousto-optic distribution for laser-driven platforms, or other modality-specific fabrics. Thus, large-scale control integration is not merely an engineering optimization; it is a necessary condition for making the abstractions above economically viable.

Taken together, these principles lead to a simple conclusion: scalable quantum computer architecture should be designed as a cost-performance optimization problem across the full heterogeneous system, not as an isolated exercise in maximizing qubit count or similar metrics. A practical QPU will resemble other specialized accelerators in the cloud: it must expose clean logical abstractions to compilers and applications, hide the overwhelming control, readout, and QEC traffic below the ISA boundary, and rely on hardened local hardware wherever bandwidth, latency, or energy make general-purpose processing uneconomical. Functional specialization then becomes the mechanism by which the system remains affordable: compute qubits, memories, magic-state factories, communication interfaces, loaders, and future arithmetic units should not all be forced into the same generic implementation but should instead be optimized around their distinct roles.

At the same time, abstraction placement and physical locality determine whether those units can actually be built, because moving information across chip, package, and cryogenic boundaries consumes wiring, power, and bandwidth that cannot be treated as secondary details. The absence of a favorable Rent's-rule regime is a major reason why this becomes the defining systems challenge at scale. Without strong multiplexing and local integration, the number of control and measurement connections tends to grow linearly with qubit count, driving complexity, thermal load, and cost. Of the two paths, control is generally the harder one. For most qubit modalities, the machine must deliver individually calibrated analog control signals—often with qubit-specific pulse generation—while avoiding crosstalk, excess dissipation, and noise. Readout, by contrast, often looks somewhat easier once multiplexing is introduced, allowing much of the demanding digital processing to occur in compute racks away from the qubits.

The final design tradeoff is ultimately the same one that governs classical HPC and cloud accelerators: modularity, slower clocks, or stronger separation may be acceptable only if they are compensated by lower cost, higher parallelism, or shorter time to solution elsewhere in the system. In first order, the feasible design space is governed by cost per module, number of modules, and effective clock rate; their combination must deliver a roughly constant and economically meaningful cost per useful quantum operation. This is the central message of the paper: quantum computers will be technologically unusual, but architecturally they should be engineered using familiar systems principles—specialization, abstraction, locality, integration, and cost-balanced performance.

---